# SHARK-NIR, the coronagraphic camera for LBT, moving toward construction

Jacopo Farinato*[a,i], Francesca Bacciotti[b], Carlo Baffa[b], Andrea Baruffolo[a], Maria Bergomi[a,i], Andrea Bianco[c], Angela Bongiorno[d], Luca Carbonaro[b,i], Elena Carolo[a,i], Alexis Carlotti[e], Simonetta Chinellato[a,i], Laird Close[f], Marco De Pascale[a], Marco Dima[a,i], Valentina D'Orazi[a], Simone Esposito[b,i], Daniela Fantinel[a], Giancarlo Farisato[a,i], Wolgang Gaessler[g], Emanuele Giallongo[d,i], Davide Greggio[a,i], Olivier Guyon[f], Philip Hinz[f], Luigi Lessio[a,i], Franco Lisi[b], Demetrio Magrin[a,i], Luca Marafatto[a,i], Dino Mesa[a], Lars Mohr[g], Manny Montoya[f], Fernando Pedichini[d,i], Enrico Pinna[b,i], Alfio Puglisi[b,i], Roberto Ragazzoni[a,i], Bernardo Salasnich[a], Marco Stangalini[d,i], Daniele Vassallo[a,h,i], Christophe Verinaud[e], Valentina Viotto[a,i], Alessio Zanutta[c]

[a]INAF Padova, Vicolo dell'osservatorio 5, 35122, Padova, Italy;
[b]INAF Arcetri, Largo Enrico Fermi, 5, 50125 Firenze, Italy;
[c]INAF Brera, Via Brera, 28, 20121 Milano MI
[d]INAF Roma, Via di Frascati 33, 00040 Monte Porzio Catone, Italy;
[e]Institut de Planétologie et d'Astrophysique de Grenoble, 414, Rue de la Piscine, Domaine Universitaire, 38400 St-Martin d'Hères (France);
[f]University of Arizona, Department of Astronomy/Steward Observatory, 933 North Cherry Avenue, Tucson, AZ, USA;
[g]Max Planck Institute for Astronomy, Königstuhl 17, 69117 Heidelberg, Germany;
[h]Università di Padova, Dipartimento di Fisica e Astronomia, Vicolo dell'Osservatorio 3, 35122, Padova, Italy;
[i]ADONI, Laboratorio Nazionale di Ottica Adattiva Italiano

## ABSTRACT

SHARK-NIR is one of the two coronagraphic instruments proposed for the Large Binocular Telescope. Together with SHARK-VIS (performing coronagraphic imaging in the visible domain), it will offer the possibility to do binocular observations combining direct imaging, coronagraphic imaging and coronagraphic low resolution spectroscopy in a wide wavelength domain, going from 0.5μm to 1.7μm. Additionally, the contemporary usage of LMIRCam, the coronagraphic LBTI NIR camera, working from K to L band, will extend even more the covered wavelength range.

In January 2017 SHARK-NIR underwent a successful final design review, which endorsed the instrument for construction and future implementation at LBT. We report here the final design of the instrument, which foresees two intermediate pupil planes and three focal planes to accomodate a certain number of coronagraphic techniques, selected to maximize the instrument contrast at various distances from the star.

Exo-Planets search and characterization has been the science case driving the instrument design, but the SOUL upgrade of the LBT AO will increase the instrument performance in the faint end regime, allowing to do galactic (jets and disks) and extra-galactic (AGN and QSO) science on a relatively wide sample of targets, normally not reachable in other similar facilities.

## 1. INTRODUCTION

SHARK (System for coronagraphy with High order Adaptive optics from R to K band) [1] is an instrument proposed for the LBT[2] in the framework of the "2014 Call for Proposals for Instrument Upgrades and New Instruments". It is composed by two channels, covering different and partially overlapping wavelength ranges: SHARK-VIS[3] will work from 0.5μm to 1μm, while SHARK-NIR[4] is dedicated to near infrared bands, Y, J and H, operating from 0.96μm to 1.7μm. SHARK-NIR successfully passed the Final Design Review (FDR), and have been approved to undergo the construction phase, to be later installed at LBT. SHARK will exploit, in its binocular fashion, unique challenging science ranging from exoplanet search and characterization to star forming regions with simultaneous spectral coverage from R to H band, taking advantage of the excellent performances of the LBT AO[5] systems, based on the Pyramid Wave Front Sensor[6] (PWFS) and on the Adaptive Secondary Mirrors[7].

The spectral coverage will become even larger when used in combination with LMIRcam of LBTI[8], which will be upgraded soon to work in K band, and will thus offer coronagraphic direct imaging from K to M band. In this scenario, LBT will have the possibility to make contemporary coronagraphic observations with three instruments:

- SHARK-NIR on one arm, operating between Y and H bands
- SHARK-VIS and LMIRcam on the other arm, that will operate contemporary (through a dedicated dichroic splitting the visible from the infrared light), the first in V,R I and Z bands, the second in K, L and M bands which is a unique scenario for coronagraphy in the framework of the modern planet finders. Another unique characteristic of LBT is that, with the foreseen upgrade of the AO system (SOUL, see [9]), the performance of the AO will be pushed in two different directions:
- in the eXtreme Adaptive Optics (XAO) regime, by upgrading the ASM speed to 2KHz and improving the controller in order to increase the number of possible corrected modes from the current 400 to about 600
- in the faint end regime, by gaining between 1 and 2 magnitude using a new detector with nearly 0 Read out Noise (RoN), increasing a lot the sample of possible exo-planetary systems to be exploited, and allowing to make also extra-galactic science by characterizing for example the morphology of faint targets such as AGN and QUASARS.

This paper describes the SHARK-NIR instrument in its final configuration, which has been recently approved to move toward construction by the LBTO board.

## 2. SHARK-NIR SCIENCE CASE

The direct detection of extrasolar planets is one of the most exciting goals of SHARK-NIR. In fact, the resolution achievable with a 10-m class telescope allows to access, in the NIR domain, gaseous giant planets of Jupiter size or bigger, which still is a very challenging task to be achieved, due to the very high contrast and vicinity to the hosting star required. There are several scientific goals to be possibly exploited in the exo-planet science case, ranging from the direct detection of unknown giant planets, to the follow up of known planets (through spectroscopic and photometric characterization).They require of course the implementation of a spectroscopic mode with modest spectral resolution that is currently foreseen in SHARK-NIR through a long slit positioned into the intermediate focal plane.

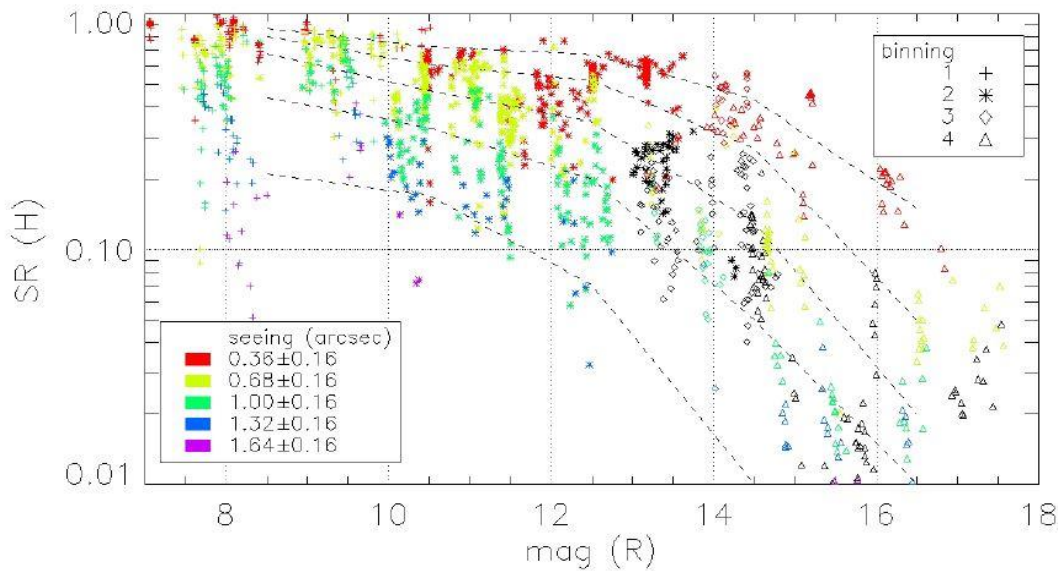

*Figure 1: a summary of the FLAO performance obtained in H band with different observing conditions*

But the science to be exploited with SHARK-NIR is definitely not only limited to the exo-planets case. In fact, the study of proto-planetary disks is fundamental to comprehend the formation of our own solar system as well as of extrasolar planetary systems. To understand how matter aggregates to form the building blocks of planetary bodies, there is the need to investigate not only the evolution of the disk itself, but also the role of jets in shaping its structure. This requires observing the system at high angular resolution as close as possible to the parent star, occulting its light to enhance the area where the interplay between the accretion and ejection of matter dominates the dynamics.

Other very interesting and challenging topics can be found in the extragalactic science, where the capabilities of SHARK-NIR in terms of spatial resolution and contrast enhancement may be applied to study the AGN-host relations as well as Dumped Ly-α systems (DLAs), to constrain the Black Hole feeding mechanism and to trace, in bright quasars, molecular outflows powerful enough to clean the inner kilo-parsec and quench the star formation.

There is anyhow an important feature of the LBT AO which, exploited in the proper way, may give to SHARK-NIR the possibility to explore unique coronagraphic science. The Pyramid WFS has a demonstrated gain in sensitivity compared to other WFSs commonly used, such as the Shack-Hartmann ([10], [11], [12], [13], [14]). This fact gives to the LBT AO systems the capability to achieve high SR (of the order of 70%) at moderately faint magnitude (R~12 or even occasionally fainter, depending on the observing conditions), as it is shown in the impressive collection of FLAO results reported in Figure 1.

This excellent performance will be further enhanced with the implementation of the AO upgrade SOUL (**S**ingle Conjugated Adaptive **O**ptics **U**pgrade for **L**BT), as it is shown in Figure 2, where, depending on the CCD choice, the capability of achieving SRs as high as 70% can be pushed to star magnitudes up to 13.5, and it has to be emphasized that these curves have been computed in non-excellent seeing conditions (0.8").

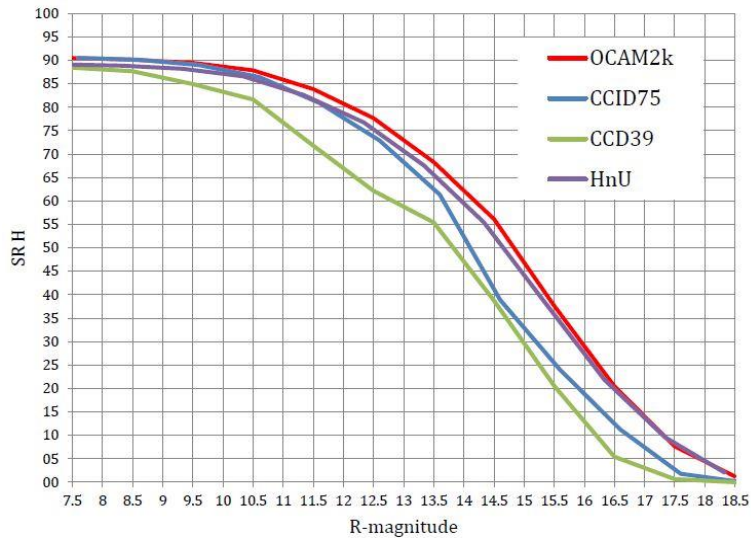

*Figure 2: the SOUL improvement in the LBT AO performance (the current is in green line) with different CCD choices, computed with a seeing of 0.8"*

This performance will open the field of high-contrast AO coronagraphic imaging to stars much fainter than required by other coronagraphic instruments, allowing deep search for planets around targets like, e.g., M dwarfs in nearby young associations and solar type stars in nearby star-forming regions (Taurus-Auriga at 140 pc). Also in the extragalactic field, the sample of AGN and, above all, of Quasars to be explored will go from a few tenths to a few hundreds, changing the perspective of the science to be achieved. This is definitely the characteristic that may give to SHARK-NIR unique opportunities in the coronagraphic instrument scenario.

## 3. SHARK AT THE TELESCOPE

SHARK will be installed at the entrance foci of LBTI (LBT Interferometer), as it is shown in Figure 3, using two deployable dichroics to feed the two SHARK channels. In this way, on the VIS side, the IR light is totally transmitted to LBTI, while on the NIR side, the NIR light will be sent to SHARK-NIR. The dichroics may be positioned just before the entrance window of LBTI, the latter transmitting the IR light to the interferometric focus and reflecting the VIS light to the Pyramid WFS. Such a dichroic, on the VIS channel would pick-up only a certain amount (selectable) of the VIS light, to feed with the rest the WFS, while on the NIR channel would pick up only the Y, J and H bands, letting all the visible light going through. With this setup, SHARK will provide possible contemporary observations from R to K bands. Such a flexible configuration with several combined binocular observing modes is reflecting the request coming from the principal science cases, for which simultaneous observations in the VIS and NIR domain are required. In the following we show the conceptual opto-mechanical study of the NIR channel.

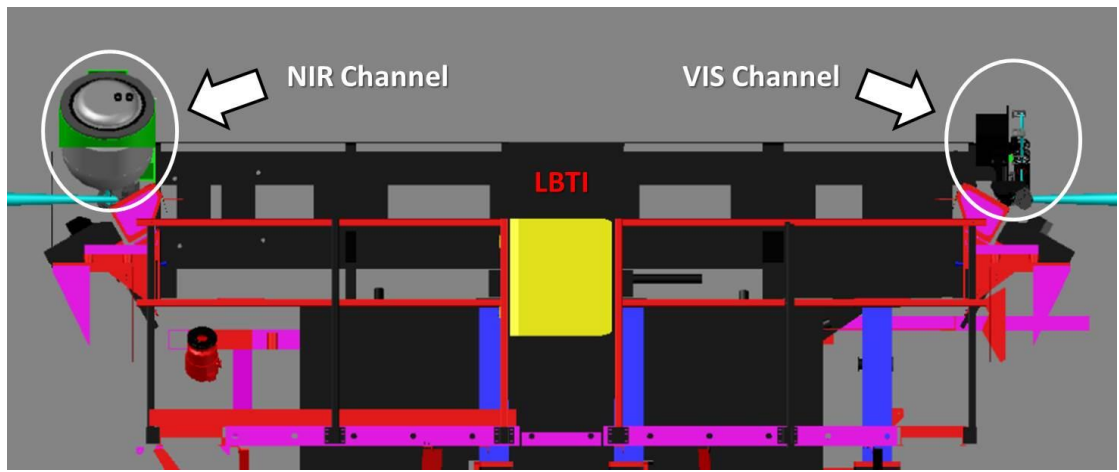

*Figure 3: a possible installation of the two SHARK channels at the LBTI entrance foci*

# 4. SHARK-NIR INSTRUMENT DESCRIPTION

The basic idea of SHARK-NIR is a camera for direct imaging coronagraphy and spectroscopy, using the corrected wavefront provided by the LBT Adaptive Secondary Mirror (ASM), operated through one of the existing LBTI AO WFS.

Being SHARK-NIR also a coronagraphic instrument, the camera has to be designed to accomplish an extreme performance, ideally not to decrease the correction provided by the AO system. In fact, all the coronagraphic techniques that may be implemented need a SR as high as possible to provide very good contrast. This requires optics machined to a state of the art technology and polished to nanometric level of roughness, properly aligned and installed on very robust mounts. The whole instrument mechanics has to be very stiff and designed to minimize the effects of flexures.

Additionally, to maintain the performance as good as possible at every observing altitude, it is necessary to implement an atmospheric dispersion corrector (ADC) to compensate for the atmospheric dispersion. Some of the foreseen science cases need to perform the field de-rotation, to accomplish which the whole instrument has to be mounted on a mechanical bearing.

A NIR camera, based on an Teledyne H2RG detector, cooled at about 80°K to minimize the thermal background, will provide a FoV of the order of 18"x18" operating in Y, J and H bands, with a plate scale foreseeing a bit more than two pixels on the diffraction limit PSF at 0.96μm.

A few subsystems have been introduced in the instrument design with the purpose of optimizing the instrument performance.

For the non-common path aberrations (NCPA) minimization, a local DM (ALPAO DM 97-15) has been introduced into the first pupil plane, allowing a local removal of the aberrations. The same DM, used in Tip-Tilt (T-T) fashion, may be used to correct undesired PSF movements during a scientific exposure. The latter correction requires a dedicated T-T sensor (based on the First Light C-RED2 camera), which has been placed after the first pupil plane, into the collimated beam (a beam splitter will pick-up few percent of the light and will send it to the sensor). A Wave Front Computer (WFC, which will be realized by Microgate) will allow the fast T-T correction achieved with the DM, which will at the same time maintain the proper shape for the Non Common Path Aberrations (NCPA) local compensation.

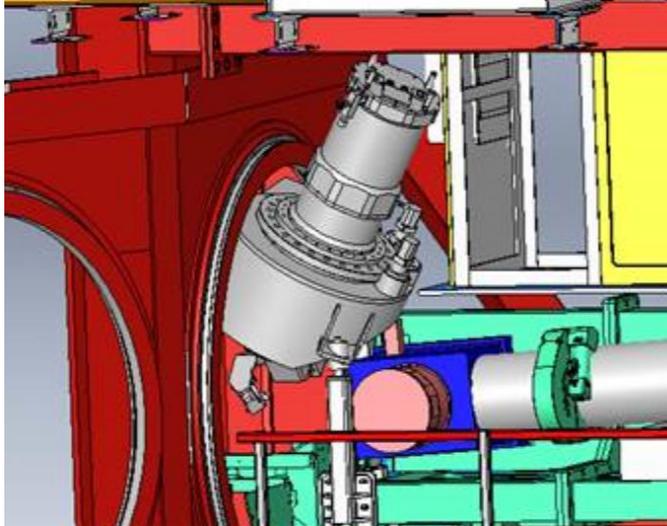

*Figure 4: the SHARK-NIR instrument positioned at the entrance of LBTI*

Between the DM and the beam splitter feeding the T-T sensor, a filter wheel positioned at 50mm from the pupil plane carries the apodizing masks. These kinds of masks are normally placed exactly into the pupil plane, which is occupied by the DM in our design. We have carefully evaluated the impact of having the masks slightly displaced with respect to the pupil plane, and it turned out that the effect is basically negligible with the considered coronagraphic techniques if the masks are designed to take this fact into account.

As we mentioned at the beginning of this section, SHARK-NIR has essentially three observing modes:
- The **Direct Imaging mode**, in which SHARK-NIR will provide an unobstructed FoV of 18"x18", with a correction which is nominally nearly perfect over the full FoV. Even with the ADC inserted (which is deployable, to have the best possible optical quality when observing at small zenithal angles), the optical performance remains very good, ensuring for example, at a zenithal distance of 50°, an on-axis SR >96%, while at the detector corner (at the edge of the field diagonal) it decreases to ~92%.
- The **Coronagraphic mode**, that will give the possibility to exploit coronagraphic imaging using a variety of possible techniques, which have the purpose to dim (ideally cancel) the light of the central star, in a way to enhance the contrast in the vicinity of the star itself. This will allow to detect much fainter companions (exo-planets case for example) or to explore the morphology of the object under study (Jets/Disks case and AGN/QSO case). They are characterized by different operating distances from the central star (Inner Working Angle, IWA), different contrast that can be reached at a given distance, different throughput and various FoVs.

A few of these techniques will be implemented, in a way to fulfill as much as possible the different needs of the different science cases, and the baseline is to provide:
- o Gaussian Lyot, which requires a gaussian stop into the 1st focal plane and a pupil stop on the 2nd pupil plane
- o Shaped Pupil, which requires an apodizing mask into the 1st pupil plane and an occulting mask into the 1st focal plane
- o APLC, which requires an apodizing mask into the 1st pupil plane, an occulting mask into the 1st focal plane and a pupil stop into the 2nd pupil plane

Additional simulations are currently ongoing to evaluate also other very interesting techniques, such as the APP, the Vortex and the 4 quadrant.
- The **Spectroscopic mode**: A long-slit spectroscopic (LSS) coronagraphic mode will be implemented in SHARK, with two different resolutions: a low-resolution mode (R~ 100, realized through a prism), in order to target faint targets, and a medium-resolution mode (R~700, realized through a grism) to get spectral information of the faint objects around the bright targets.

## 5. THE OPTO-MECHANICAL CONCEPT

In this section we give an overview of the opto-mechanical design of SHARK-NIR, showing the main subsystems of the instrument, and recalling its main characteristics. The instrument is designed to operate in the wavelength range going from 1μm to 1.7μm.

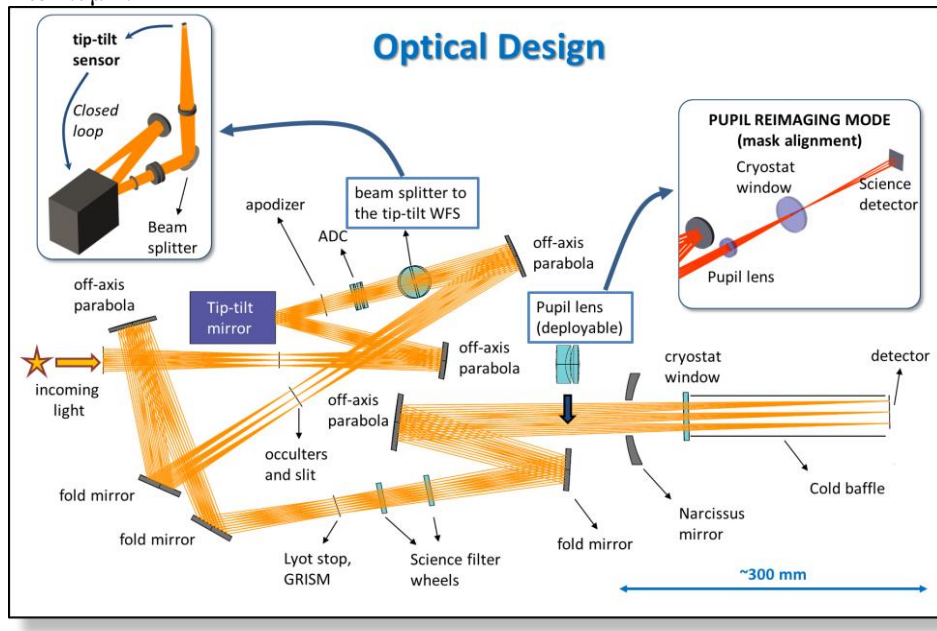

*Figure 5: The Optical design of SHARK-NIR*

The optical design is shown in Figure 5, the full circular FoV is of the order of 15" (~21.5" on the diagonal), and the main characteristics are the following:
- An off-axis parabola (OAP-1) is creating a pupil plane of about 12mm of diameter onto the DM. A possible local DM to be used has already been identified, the ALPAO 97-15, characterized by a maximum pupil size of 13.5mm in diameter, by 97 actuators (normally the NCPA correction is limited to the first 15-20 modes), a large PtV stroke of the actuators of about 40-60μm and a bandwidth of about 1250Hz.
- A filter wheel will select between different apodizing masks, positioned 50mm after the pupil plane.
- Immediately after, in the collimated beam, the ADC is placed.
- The ADC is deployable, in a way to optimize the system performance at observing altitudes that do not require the correction (normally for zenithal distances smaller than 25°-30°).
- Between the ADC and the second off-axis parabola (OAP-2), a beam splitter is placed to send a small portion of the light (~5%) to a very simple tip-tilt sensor (which is placed in vertical position with respect to the plane of the drawing), composed of a lens and a commercial camera (First Light C-RED2) sensitive to Y, J and H bands. The T-T sensor gives the advantage to monitor (at high frequency, by windowing the camera can go as fast as ~15KHz) possible drifts of the spot during a single exposure, to be then compensated with the local DM, ensuring in this way to maintain the proper mask alignment and moreover to minimize the residual jitter down to a few mas level.
- OAP-2 is refocusing the beam on an intermediate focal plane (FP-1), where a filter wheel can select between different occulting masks (10 positions are foreseen). The same wheel accommodates a couple of low spectral

- resolution grism (R~100 and R~1000) to perform spectral characterization of the science targets.
- After a folding mirror, a third off-axis parabola (OAP-3) is creating the 2$^{nd}$ re-imaged pupil plane, where a filter wheel (9 positions foreseen) can select between different pupil stops used to properly mask the spiders and the secondary mirror, to minimize diffraction effects. On the same collimated beam, two additional filter wheels (positioned between the pupil plane and a folding mirror, Fold-3) will allow the insertion of seven scientific filters each. They both have eight positions.
- After a folding mirror, the fourth off-axis parabola (OAP-4) is creating the final focal plane onto the detector, where the diffraction limit PSF is Nyquist sampled at 1μm. A deployable small optical group can be inserted between OAP-4 and the cryostat window, with the purpose to create an image of the pupil onto the detector, which can be used before each scientific exposure to properly calibrate and compensate pupil shifts.

The whole bench is installed on a mechanical bearing, allowing the field rotation whenever required from the science cases.

The entrance window of the camera dewar is kept at 200mm of distance from the detector, being 180mm the minimum length of the baffle which has to be implemented in front of the camera to minimize the thermal background.

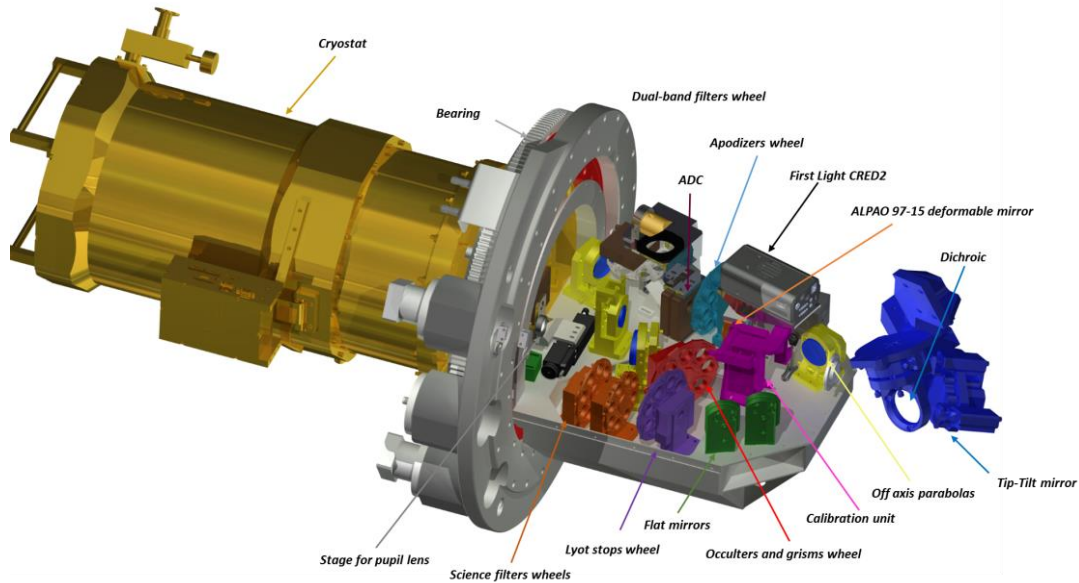

*Figure 6: the opto-mechanical concept of the SHARK-NIR optical bench*

The camera is Cryogenic, and will be using an HAWAII II detector. The LN2 tank shall ensure a hold time of about 28 hours.

## CONCLUSIONS

SHARK-NIR is an instrument designed for high contrast imaging that will exploit the extreme adaptive optics correction provided by the Pyramid based LBT AO systems. As described, three observing modes will be implemented: direct imaging, coronagraphic imaging and spectroscopy. This flexibility will allow to take full advantage of the LBT binocular capabilities and will give clear opportunities to obtain primary science goals, in combination with the excellent LBT AO performance and also in synergy with other LBT instruments designed for very high contrast imaging, such as SHARK-VIS and LMIRCam.

Furthermore, the SOUL upgrade will improve even more the performance of the AO systems, increasing the speed of the adaptive secondary mirrors and increasing the WFSs sensitivity using L3-CCDs. As shown in Section 2, we expect 1-2 magnitude fainter targets to be reachable with respect to other similar instruments.

The impact on the SHARK-NIR science would then be clear, allowing LBT to observe, in the NIR domain, many more nearby small mass stars and solar type stars in nearby star-forming regions (e.g. Taurus at 140 pc), complementing somehow the surveys of SPHERE[15], GPI[16], MAGAO[17] and SCExAO[18].

## REFERENCES

[1] Farinato, J.; Pedichini, F.; Pinna, E.; Baciotti, F.; Baffa, C.; Baruffolo, A.; Bergomi, M.; Bruno, P.; Cappellaro, E.; Carbonaro, L.; Carlotti, A.; Centrone, M.; Close, L.; Codona, J.; Desidera, S.; Dima, M.; Esposito, S.;

Fantinel, D.; Farisato, G.; Fontana, A.; Gaessler, W.; Giallongo, E.; Gratton, R.; Greggio, D.; Guerra, J. C.; Guyon, O.; Hinz, P.; Leone, F.; Lisi, F.; Magrin, D.; Marafatto, L.; Munari, M.; Pagano, I.; Puglisi, A.; Ragazzoni, R.; Salasnich, B.; Sani, E.; Scuderi, S.; Stangalini, M.; Testa, V.; Verinaud, C.; Viotto, V.; "SHARK (System for coronagraphy with High order Adaptive optics from R to K band): a proposal for the LBT 2nd generation instrumentation", SPIE Proc. 9147, id. 91477J 10 pp. (2014).

[2] Hill, J. M. and Salinari, P., "Large Binocular Telescope project," Proc. SPIE, 4004, 36–46 (2000).

[3] Stangalini, M.; Pedichini, F.; Centrone, Mauro; Esposito, S.; Farinato, J.; Giallongo, E.; Quirós-Pacheco, F.; Pinna, E.; "The solar system at 10 parsec: exploiting the ExAO of LBT in the visual wavelengths", SPIE Proc. 9147, id. 91478F 7 pp. (2014).

[4] Farinato, J., Baffa, C., Baruffolo, A., Bergomi, M., Carbonaro, L., Carlotti, A., Centrone, M., Codona, J., Dima, M., Esposito, S., Fantinel, D., Farisato, G., Gaessler, W., Giallongo, E., Greggio, D., Hink, P., Lisi, F., Magrin, D., Marafatto, L., Pedichini, F., Pinna, E., Puglisi, A., Ragazzoni, R., Salasnich, B., Stangalini, M., Verinaud, C. and Viotto, V., "The NIR arm of SHARK (System for coronagraphy with High order Adaptive optics from R to K band)", International Journal of Astrobiology, 14(3), 365-373 (2015)

[5] Esposito, S., Riccardi, A., Pinna, E., Puglisi, A. T., Quirós-Pacheco, F., Arcidiacono, C., Xompero, M., Briguglio, R., Busoni, L., Fini, L., Argomedo, J., Gherardi, A., Agapito, G., Brusa, G., Miller, D. L., Guerra Ramon, J. C., Boutsia, K. and Stefanini, P. "Natural guide star adaptive optics systems at LBT: FLAO commissioning and science operations status", Proc. SPIE, 8447, 84470U (2012)

[6] Ragazzoni, R., "Pupil plane wavefront sensing with an oscillating prism," Journal of Modern Optics 43, 289 (1996).

[7] Riccardi, A.; Xompero, M.; Briguglio, R.; Quirós-Pacheco, F.; Busoni, L.; Fini, L.; Puglisi, A.; Esposito, S.; Arcidiacono, C.; Pinna, E.; Ranfagni, P.; Salinari, P.; Brusa, G.; Demers, R.; Biasi, R.; Gallieni, D.; "The adaptive secondary mirror for the Large Binocular Telescope: optical acceptance test and preliminary on-sky commissioning results", SPIE 7736, 79 (2010)

[8] Hinz, P.M., Angel, J.R.P., McCarthy, D.W., Hoffman, W.F., and Peng, C.Y., "The large binocular telescope interferometer," in Proc. SPIE, 4838, 108 (2003).

[9] Pinna, E., Pedichini, F., Farinato, J., Esposito, S., Centrone, M., Puglisi, A., Carbonaro, L., Agapito, G., Riccardi, A., Xompero, M., Hinz, P., Montoya, M., and Bailey, V., "XAO at LBT: current performances in the visible and upcoming upgrade," Proc. of AO4ELT4 conference.

[10] Ragazzoni, R.; Farinato, J.; "Sensitivity of a pyramidic Wave Front sensor in closed loop Adaptive Optics", A&A, 350, L23 (1999)

[11] Vérinaud, C.; Le Louarn, M.; Korkiakoski, V.; Carbillet, M.; "Adaptive optics for high-contrast imaging: pyramid sensor versus spatially filtered Shack-Hartmann sensor", MNRAS, 357, L26 (2005)

[12] Esposito, S.; Riccardi, A.; "Pyramid Wavefront Sensor behavior in partial correction Adaptive Optic systems", A&A, 369, L9 (2001)

[13] Costa, Joana B.; Feldt, Markus; Wagner, Karl; Bizenberger, Peter; Hippler, Stefan; Baumeister, Harald; Stumpf, Micaela; Ragazzoni, Roberto; Esposito, Simone; Henning, Thomas; "Status report of PYRAMIR: a near-infrared pyramid wavefront sensor for ALFA", SPIE Proc. 5490, 1189 (2004)

[14] Ghedina, Adriano; Cecconi, Massimo; Ragazzoni, Roberto; Farinato, Jacopo; Baruffolo, Andrea; Crimi, Giuseppe; Diolaiti, Emiliano; Esposito, Simone; Fini, Luca; Ghigo, Mauro; Marchetti, Enrico; Niero, Tiziano; Puglisi, Alfio; "On Sky Test of the Pyramid Wavefront Sensor", SPIE Proc. 4839, 869 (2003)

[15] Beuzit, J.-L.; Boccaletti, A.; Feldt, M.; Dohlen, K.; Mouillet, D.; Puget, P.; Wildi, F.; Abe, L.; Antichi, J.; Baruffolo, A.; Baudoz, P.; Carbillet, M.; Charton, J.; Claudi, R.; Desidera, S.; Downing, M.; Fabron, C.; Feautrier, P.; Fedrigo, E.; Fusco, T.; Gach, J.-L.; Giro, E.; Gratton, R.; Henning, T.; Hubin, N.; Joos, F.; Kasper, M.; Lagrange, A.-M.; Langlois, M.; Lenzen, R.; Moutou, C.; Pavlov, A.; Petit, C.; Pragt, J.; Rabou, P.; Rigal, F.; Rochat, S.; Roelfsema, R.; Rousset, G.; Saisse, M.; Schmid, H.-M.; Stadler, E.; Thalmann, C.; Turatto, M.; Udry, S.; Vakili, F.; Vigan, A.; Waters, R.; "Direct Detection of Giant Extrasolar Planets with SPHERE on the VLT", ASP Proc. 430, 231 (2010)

[16] Macintosh, B. A., Graham, J. R., Palmer, D. W., Doyon, R., Dunn, J., Gavel, D. T., Larkin, J., Oppenheimer, B., Saddlemyer, L., Sivaramakrishnan, A., Wallace, J. K., Bauman, B., Erickson, D. A., Marois, C., Poyneer, L. A., and Soummer, R., "The Gemini Planet Imager: from science to design to construction," Proc. SPIE 7015, 31 (2008).

[17] Close, L. M.; Males, J. R.; Kopon, D. A.; Gasho, V.; Follette, K. B.; Hinz, P.; Morzinski, K.; Uomoto, A.; Hare, T.; Riccardi, A.; Esposito, S.; Puglisi, A.; Pinna, E.; Busoni, L.; Arcidiacono, C.; Xompero, M.; Briguglio, R.; Quiros-Pacheco, F.; Argomedo, J., "First closed-loop visible AO test results for the advanced adaptive secondary AO system for the Magellan Telescope: MagAO's performance and status", SPIE Proc. 8447, 0 (2012)

[18] Guyon, O.; Martinache, F.; Clergeon, C.; Russell, R.; Groff, T.; Garrel, V.; "Wavefront control with the Subaru Coronagraphic Extreme Adaptive Optics (SCExAO) system", SPIE Proc. 8149, 894293 (2011)